**Using the Spatial Clustering Approach with Constraints in Automation of Pavement Project Priority and Management Methods**

Shadi Hanandeh[1], Omar Elbagalati [2], Mustafa Hajij[3]

[1]Associate Professor civil engineering department, Al-Balqa Applied University,
Email: hanandeh@bau.edu.jo. (Corresponding author)

[2] Michelin Mobility Intelligence Director of Market Insights and Research,
Email: Omar.elbagalati@michelin.com ,

[3] Assistant Professor, Department of Data Science, University of San Francisco. Email
mhajij@usfca.edu

## ABSTRACT

The primary focus of pavement management software solutions and research efforts since its establishment in the 1990s has predominantly revolved on three key parameters: the condition of the pavement, the financial implications of repair, and the availability of funding. Nevertheless, the examination of the importance of the geographical characteristics of the candidate projects in the decision-making procedure remains little investigated. In the majority of instances, this limitation allows for the inclusion of construction sites that are functioning simultaneously but are geographically distant in the planned monitoring and managing project schedule, hence presenting significant challenges. This study introduces a novel approach to incorporate the spatial characteristics of pavement segments into the analysis as part of the prioritization process for maintenance and rehabilitation (M&R). The recently devised methodology aims to integrate pavement segments with converging geographical coordinates, with the objective of restoring them during the same fiscal year while adhering to the predetermined budgetary constraints established for the project. An optimal combination of these variables would result in a decrease in the duration of travel for employees, materials, and equipment between construction sites, as well as an enhancement in collaboration and communication among the construction teams. The present study introduces a groundbreaking spatial clustering method that autonomously identifies projects that align with a specified budget and geographical limitation definition. The newly developed method was validated using a total of 1,800 projects derived from two real-world scenarios, namely the City of Milton, Georgia, and the City of Tyler, Texas.

## INTRODUCTION

One major responsibility of the transportation agencies is to manage the use of available funds to achieve adequate overall pavement network condition [1]. Considering the rapid deterioration of pavements, due to the continuous increment in the traffic volume, the task of investigating the most cost-effective strategy in managing the pavement network has become very challenging [2] . Therefore, several agencies hire consultants, pavement engineers, and qualified inspectors to provide the decision makers with an accurate evaluation of the pavement network condition, a list of candidate maintenance and rehabilitation (M&R) projects, and a detailed rehabilitation projects' schedule [3]. Accordingly, several pavement management systems (PMS) software products have been developed to assist the pavement engineers in the M&R prioritization, such as PAVER, Cartegraph, PERS, StreetSaver and PAVEMAN [4]. In addition, several optimization algorithms for M&R prioritization have been recently introduced (Babashamsi et al., 2016, Gowda et al., 2015, [7], [8],[9],[10],[11]. However, up to the authors' knowledge, no pavement software product or optimization algorithm has taken into consideration the spatial correlations between the candidate M&R projects. Per this current practice, the recommended M&R projects' schedule consistently encompasses simultaneously running but spatially scattered construction sites, which are very challenging to monitor and manage.

Managing multiple projects sites is a well-known common challenge in the construction industry [12],[13]. Obviously, if these projects are spatially scattered, this would add more complexity to that challenge [14].Therefore, in most cases, the transportation agency would request a reformulation of the proposed schedule, so that M&R projects within the same geographic zone are combined and conducted in the same fascial year. This strategy allows the contractor to perform the fieldwork using continuous workflow, which provides a significant decrease in cost, timing, and mobilization of labor, equipment, and materials ([15] and [16]. Therefore, the transportation agency receives lower proposals from the bidding contractors. However, the process of reformulating the project schedule is highly subjective and time-consuming, as there is no automated algorithm to perform the task. This study introduces a novel approach to perform the reformulation process by integrating pavement segments' spatial characteristics in the (M&R) prioritization analysis.

Proposed herein, a clustering algorithm that naturally partitions the projects while considering simultaneously the budget constraints, the schedule, and spatial characteristics. The number of clusters in the proposed method is determined by the total number of years allocated for the projects. The algorithm starts by finding the centroid for the clusters, then the clusters grow by gradually increasing the number of nearest neighbors around the centroids while taking into consideration the budget and the schedule constrains. Each cluster is finalized when its allocated budget is reached. The factors that form the clusters' constraints can be given weights to determine when the points join the cluster. Points with higher weights joint the cluster before points with lower weights. In our case, higher weights are assigned to projects closer to the cluster centroid, have a small budget and have earlier schedules. The closeness to the centroid factor is given the highest priority to enter the cluster, the other factors: the budget size and project schedule are considered afterward in this order. In the presented article, the terminologies of M&R projects and pavement segments are interchangeable. In addition, the terminologies of cluster, zone, and fascial year are interchangeable.

## LITERATURE REVIEW AND TECHNICAL BACKGROUND

### Pavement Management Budget analysis and Project planning

Pavement management is a process aims at planning the M&R projects of the roadway network to achieve a satisfactory level of service over the entire network while taking into consideration the budget constraints[9]. Generally, there are two types of M&R planning problems, specifically, the budget planning problems and the budget allocation problems[17]. In the budget planning problem, a required budget needs to be determined, so that a target overall pavement network condition is achieved. In the budget allocation problem, the available budget is known, and the most cost-effective allocation of the available funds is to be determined. In both scenarios, an M&R project schedule is proposed, and funds are distributed among the different M&R Categories. Mainly, there are three pavement M&R categories, namely, major M&R, global M&R, and localized M&R [18]. The major M&R category includes heavy rehabilitation activities such as reconstruction and Hot Mix Asphalt (HMA) overlays. The global M&R category involves preventive activities such as seal coats and micro-surfacing[19],[20],[21] . The localized M&R category encompasses minor activities that aims at maintaining safe drive on the roadway surface, such as patching a pothole or sealing a crack [22]. Major and global M&R projects are the focus of the project planning and project allocation problems because of the continuous workflow pattern, and because they consume most of the available funds and construction time. On the other hand, localized maintenance activities are mostly conducted in-house, as they are very light and safety-related; therefore, they are out of the consideration of the project planners.

**Efforts in Incorporating Engineering Criteria into PMS project planning**

In this section, noteworthy studies aimed at incorporating more than the available budget and the pavement condition criteria into the PMS project planning are summarized. In 2009, Yang and co-authors introduced an algorithm for the spatial clustering of the pavement segments [2]. The authors utilized the fuzzy c-mean clustering method to divide the pavement segments selected for repair in a fascial year into clusters (projects) based on condition similarity, work continuity, and project length. However, the algorithm doesn't deal with projects recommended for repair at different fascial years. In 2010, [23] developed a model utilizing Geographic information system (GIS) for M&R project planning in Nepal [23].The model takes into consideration locations of roadside slope failure, which is critical in the pavement M&R planning in mountain regions. In 2012, Zhou and Wang, introduced a co-location based decision tree (CL-DT) for pavement management [24]. The algorithm takes into consideration the spatial distance between the pavement segments; however, the available budget constraints and the individual project cost were not considered in the model development. Almeida and co-authors developed a methodology for unpaved road maintenance project prioritization using analytic hierarchy process (AHP) [7]. The methodology takes into account the traffic volume, the climatic condition, and social parameters. The AHP-based approach results are subjective in nature, and no optimum solution can be achieved without being interpreted as subsidies to the decision-making process. In 2015, Yu and co-authors introduced a multi-objective optimization model that integrates environmental impacts to the PMS project planning [25]. Environmental elements that were considered in the analysis encompass, fuel consumption, vehicle operation cost, and global warming potential. On the other hand, the approach doesn’t account for the physical distance between the pavement segments.

**Partition-Based Spatial Clustering Algorithms**

The literature of clustering is vast and general references regarding clustering include ([26], [27], [28]. The problem that is considered in this paper can be classified as a partition-based clustering problem with some additional constraints that are specific to the introduced problem. Partition-based clustering algorithms are generally divided into two types: Partitioning Relocation Methods and Density-Based Partitioning. Partitioning relocation methods, such as k-mean [29] and k- medoids [30],[31] try to learn the clusters by iteratively reassigning points between subsets while trying to minimize a certain clustering criterion. Partition relocation methods usually require the number of clusters as an input to the algorithm.

Density-based methods algorithm, such as DBSCAN [32] , [33], [34] try to find densely connected components in the data. Unlike partitioning relocation clustering algorithms, density-based methods do not require this input but require other density-related parameters such as the local radius and number of samples within this radius [35]. For more on clustering algorithms, we refer the reader to the survey as [36] and the references therein.

While the number of clusters in our method is naturally given as a part of the problem, existing partitioning relocation methods do not apply directly to our situation where we require clustering the pavement segments while simultaneously maintain the constraints of the original assigned repairing fiscal year of the segment and not compromising the allocated budget levels. The classical setups of the existing partitioning relocation and density-based methods are usually indifferent to the nature of the points and only consider their spatial or metric setting. Our setting here requires additional consideration where each point is assigned a real number that represents the cost allocated to that point. Moreover, each cluster is assigned a positive real number that represents the total budget allocation in a fiscal year. Alternative methods such as genetic algorithm and artificial neural networks, which permit to create spatially the complicated schemes such as pavement and geotechnical application[37],[38].

In their study, Zhao et al. (2022) introduced a novel approach to assessing road maintenance priority. This approach is based on an unsupervised multidimensional performance data-driven model that incorporates complete multidimensional indicators. By doing so, the suggested model addresses the limitations associated with relying on a single decision index and the lack of flexibility observed in existing evaluation models. The proposed approach utilizes a hybrid model that combines the DBSCAN algorithm for density-based spatial clustering of applications with noise and the SVM for evaluating the rank and priority of various clusters within pavement sections. The findings validate the efficacy of the hybrid model in minimizing the need for human involvement and its potential to enhance the utilization of big data methodologies in the field of road maintenance management.

Ragusa et al. (2022) designed five new heuristic algorithms that replicate various management strategies that airport managers can implement within a given time horizon to reduce pavement maintenance costs while considering the impact on the airport's service level. The acquired numerical results demonstrated that no single strategy can be deemed the most effective in terms of cost and pavement quality.

Fawzy et al. (2023) developed a multi-criteria decision-making approach to prioritize road network pavement maintenance and rehabilitation. In two stages, the model computations identify maintenance and rehabilitation factors for six criteria: pavement quality (performance), road safety, traffic level, pavement age, road class, and construction quality. Then, using the Analytic Hierarchy Process (AHP) method, the weight of criteria that influence pavement maintenance and rehabilitation decision-making is calculated to generate an overall index for each road and arrange it asc the designed maintenance decision-making model prioritizes road network maintenance while considering the annual maintenance budget. The authorities prioritized pavement upkeep, which improved Egypt's road maintenance system. Recent improvements in computerized pavement distress identification have contributed to the development of new deep learning frameworks that aim to make finding cracks and ruts more accurate and faster. MixCrackNet uses deformed convolution, adjusted losses, and multiple scales attention structures to precisely and reliably segment pavement cracks in an extensive variety of datasets. CrackDiffusion adds a two-stage semantic classification framework that uses unsupervised as well as supervised learning to enhance crack detection better. Also, an unsupervised method for finding pavement ruts that uses structured light-based deep neural network algorithms is quite accurate and doesn't need any labeled data. All of these research show that improved deep learning architectures could make it much easier to find and analyze pavement problems, which would make maintenance and management of infrastructure more effective [42–44]. Gong et al. (2024) investigated more than 50 DL-based models for crack segmentation and produced a state-of-the-art review. They focused on how convolutional neural networks neural networks (CNNs), neural mechanisms of attention, and architectural optimizations might make models more accurate and faster. Hemdan and Al-Atroush (2025) added to these efforts by doing a comprehensive evaluation of smart road crack detection systems. They looked at both algorithmic methods and real-world hardware implementations, like UAVs and mobile sensors. Together, these studies show how AI-driven road maintenance is changing, showing both the problems and the huge possibilities of using deep learning, computer vision, and sensor data together to manage civil infrastructure [45,46].

## Methodology

### Notation and Definitions

The $k^{th}$ Euclidian space will be denoted by $R^k$. A point in $R^k$ will be denoted by x. A point cloud is a finite set of points in $R^k$. We will denote the $j^{th}$ coordinate of the point x in $R^k$ by $r_j(x)$. The Euclidian distance between two points x and y in $R^k$ will be denoted by $d(x, y)$.

## Problem Formulation

We start with a data set X that consists of n data points $\{x_1, \dots, x_n\}$ in $R^k$. To each point $x_i$ we are given a cost given by $f(x_i) > 0$. That is, we have a cost function $f: X \to R$ defined on the data X. We are interested in dividing the data D into N clusters $\{C_1, \dots, C_N\}$ such that $f(C_i) := \sum_{x \in C_i} f(x) \leq p_i$ where $p_i$ is a positive number that is assigned to the cluster $C_i$ representing the total cost of this cluster. We further add the global cost constraint: $\sum_{i=1}^{n} f(x_i) = \sum_{i=1}^{N} p_i$.

Remark: Ideally, we want $f(C_i)$ to be equal to $p_i$. However, given the spatial constraint on the points this may not be always possible. In our specific case the cluster cost $p_i$ is usually given a small tolerance value $t_i$. By considering the tolerance value the clustering constraint can be written $f(C_i) \in [p_i - t_{i,} p_i + t_i]$. For the simplicity of the discussion of the algorithm below we will not use this condition. However, it will be clear from the main algorithm how to achieve this constraint on each cluster $C_i$.

## Method

Given a list of costs $\{p_1, \dots, p_n\}$ we find the clusters $\{C_1, \dots, C_n\}$ such that $\sum_{x \text{ in } C_i} f(x) \leq p_i$ as follows. We start by choosing a point $CN_1$ from the data X and then start forming a cluster $C_1$ out of $CN_1$ by incrementally adding the closest points to $CN_1$ from X as long as the cost of total added points does not exceed the total cost assigned to the cluster. The cluster $C_1$ is then subtracted from the set X. The process is then repeated on the remaining data set $X \setminus C_1$, namely, we choose a point $CN_1$ and repeat a similar process to what we just described. We iterate this process until there are no more elements in the dataset.

There are some subtle points that we did not specify in the algorithm summary above such as the choice of the point CN. These details will be handled in later sections of the article.

The main algorithm, given below, requires the subroutine Radial_Neighbor_Clustering which we list as a separate algorithm for readability of the main algorithm.

**Algorithm 1: Radial_Neighbor_Clustering**

**Input**: A data X consisting of n data points $\{x_1, \ldots, x_n\}$ in $R^k$. A specific point CN in X. A positive number p representing the cost of the cluster.

**Output**: A cluster $C \subseteq X$ with $f(C) \leq p$.

1. If $f(CN) \geq p$ then return cluster that contains CN
2. Order the points $\{x_1, \ldots, x_n\} \setminus \{CN\}$ according to their distance with respect to the point CN. Name the new ordered list $L := \{y_1, \ldots, y_{n-1}\}$. Note that the list L has all elements of X except CN.
3. Set $TotalCost = f(z)$
4. Initiate an empty list $C = \{\ \}$
5. While ($TotalCost \leq p$ and $1 \leq i \leq n - 1$):
    a. If $TotalCost \leq p$: exist the While loop
    b. Else: Insert $y_i$ in C and update the variable: $TotalCost := TotalCost + f(y_i)$
6. Return C

The main algorithm can is given below.

Algorithm 2: **Main_Algorithm**

Input: A data X consisting of n data points $\{x_1, \ldots, x_n\}$ in $R^k$. The number of clusters N. An ordered list of positive number $\{p_1, \ldots, p_N\}$ representing the costs of the clusters.

Output: N clusters $\{C_1, \ldots, C_N\}$ such that $f(C_i) \leq p_i$ for $1 \leq i \leq N$.

1. Set the list $RemaningData$ to be $X$
2. Initiate an empty list $L = \{\ \}$.
3. For ($i = 1$ to $i = N$):
4. Let $CN_i$ be a random point in $RemaningData$
5. Set $C_i = Radial_Neighbor_Clustering(RemaningData, CN_i, p_i)$
6. Set $RemaningData \coloneqq RemaningData - C_i$
7. Insert $C$ in $L$
8. Return $L$

The flow chart of the main algorithm is given in Figure 1. Part (a) of step (3) selects the center of the cluster in the algorithm in which the subroutine $Radial_Neighbor_Clustering$ utilizes as an input to form the cluster. The centers of the clusters hence affect the overall quality of the final clustering results. Ideally, in our case we want the new cluster to be "as far as possible" from the existing clusters. For this reason, we develop a better method to choose the center in the next section that satisfies exactly this criterion.

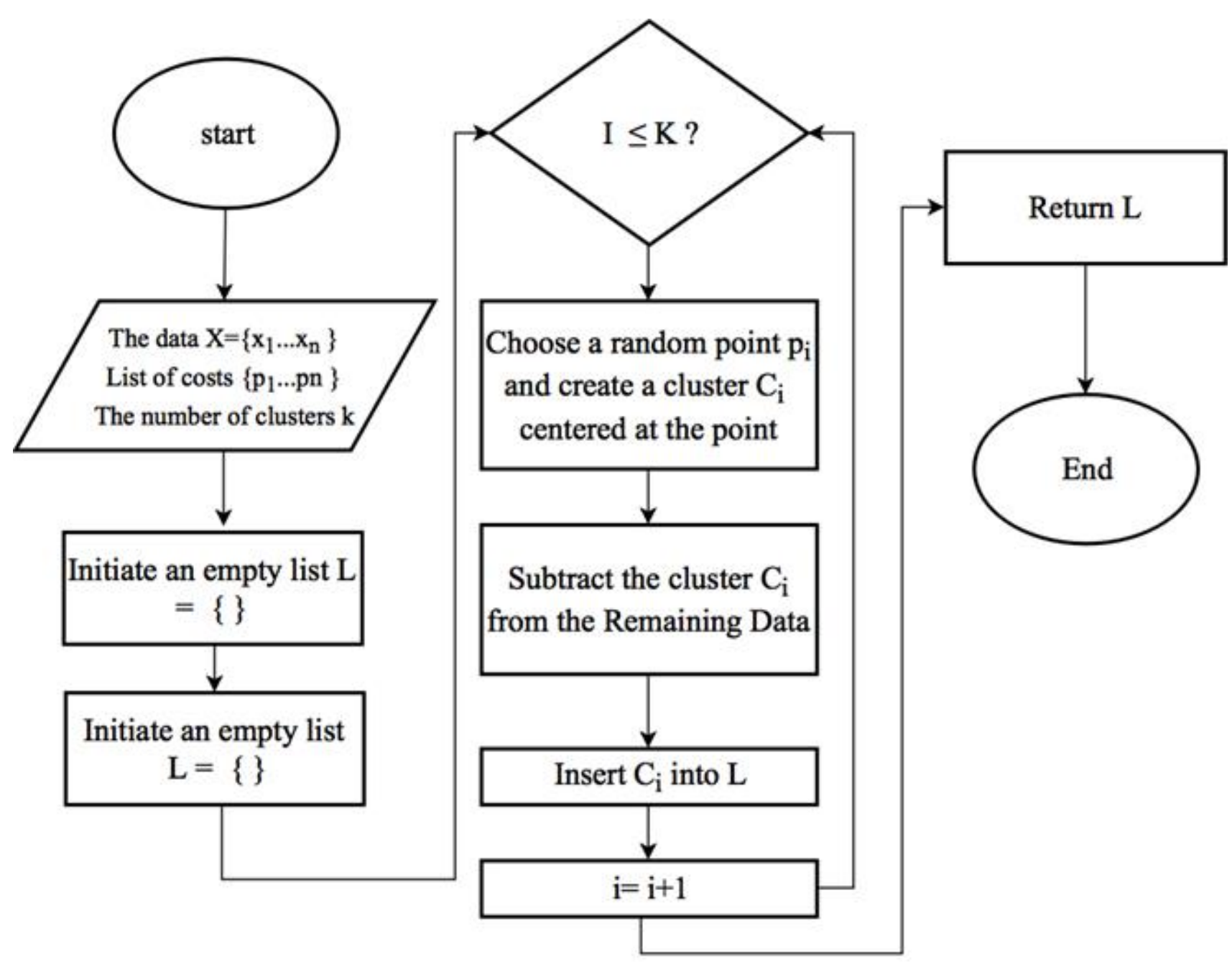

Figure 1: The flow chart of the main algorithm

**Choosing the Centers of the Clusters**

Our initial experimentation on the Main Algorithm showed that selecting the centers of the clusters (part a of step 3) is crucial to the final overall quality of the clustering results. For our particular study, we need every new cluster formed is preferred to be as far as possible from the already established clusters. For this reason, we developed the subroutine $Furthest_Point_From_Cluster$ given below. We need to define the notion of distance between a point and a cluster before we introduce this subroutine.

Let $x$ be a point in $R^k$ and $S$ be a point cloud in $R^k$. We define the distance between the point $x$ and the set $S$, denoted by $d(S,x)$ to be $\min_{s \in S} d(s,x)$. Note that if $s \in S$ then $d(S,x) = 0$. There are other methods to choose the distance between a point and a cluster and more generally one can find measure the distance between any two clusters. (Shalizi 2009) We choose this definition for its simplicity.

**Algorithm 3: Furthest_Point_From_Cluster**

Input: An ordered pair $(X, S)$ of point clouds.

Output: The furthest point $z$ in $X$ from S

1. Set $z$ to be any point in $X$.
2. Set $MaxDistance = d(S,z)$
3. For every point $y$ in $X$ do:
    a. If $d(S,y) > MaxDistance$:
        i. $MaxDistance := d(S,y)$
        ii. $z := y$

4. Return $z$

The flow chart of the furthest point from cluster algorithm is given in Figure 2. Note that the order of the input point cloud $X$ and $S$ in Algorithm is important. In other words, $Furthest_Point_From_Cluster(X, S)$ is not equal in general to $Furthest_Point_From_Cluster(S, X)$.

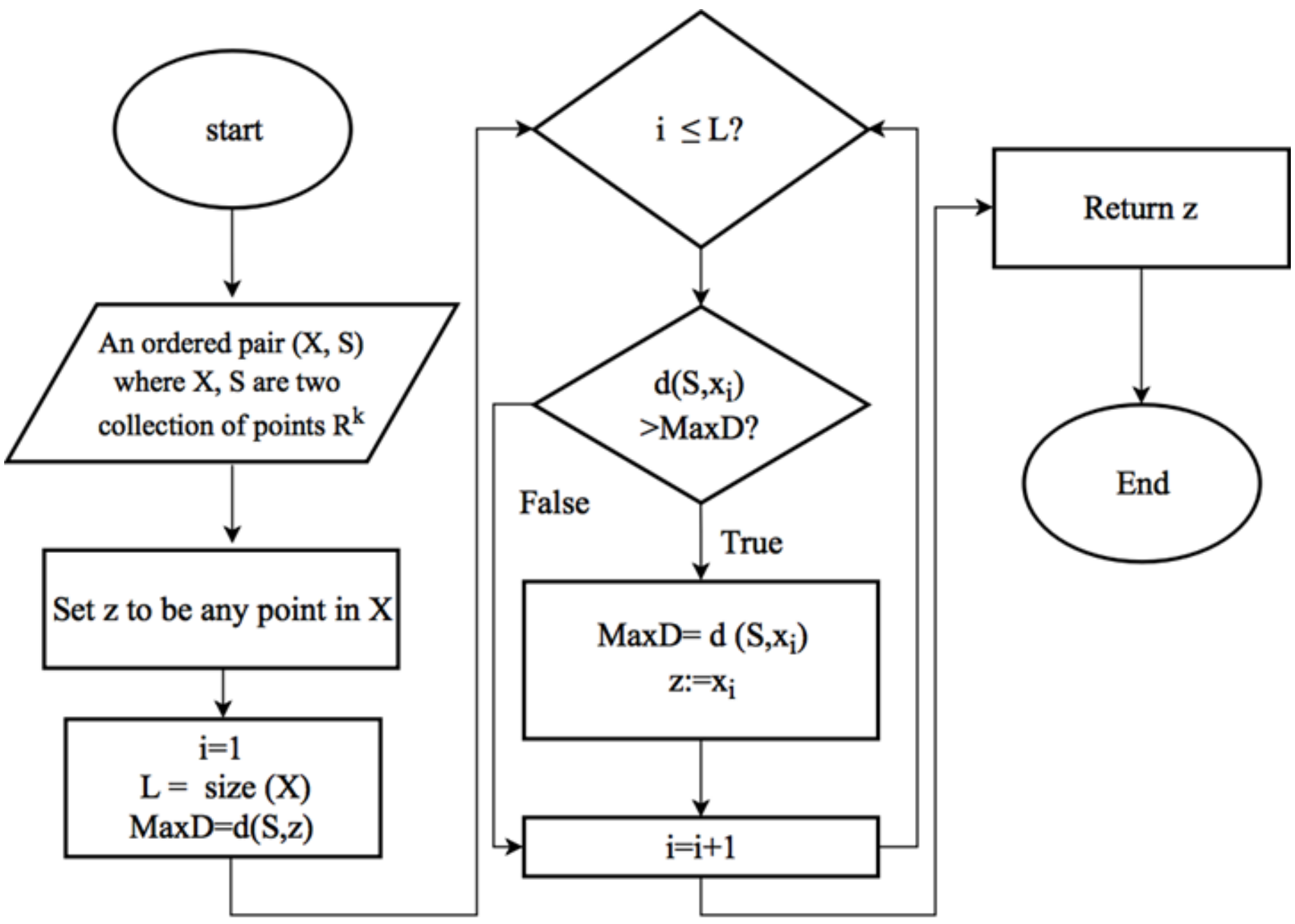

Figure 2: The flow chart of the ***Furthest_Point_From_Cluster*** algorithm

**Algorithm 4: Landark_Based_Radial_Clustering**

Input: A data $X$ consisting of n data points $\{x_1, \ldots, x_n\}$ in $R^k$. The number of clusters $N$. An ordered list of positive number $\{p_1, \ldots, p_N\}$ representing the costs of the clusters.

Output: N clusters $\{C_1, \ldots, C_N\}$ such that $f(C_i) \leq p_i$ for $1 \leq i \leq N$.

1. Set the list $RemaningData$ to be $X$.
2. Let $InitialCenter$ be the point in X with $r_i(InitialCenter) = \max_{x \in X} r_i(x)$ for some $1 \leq i \leq k$.

3. Set $center := InitialCenter$

4. Initiate an empty list $L = \{\}$.

5. For ($i = 1$ to $i = N$):

   a. Set $C_i = Radial_Neighbor_Clustering(RemaningData, center, p_i)$

   b. Set $RemaningData := RemaningData - C_i$

   c. Insert $C_i$ in $L$

   d. Set $center := Furthest_Point_From_Cluster(RemaningData, X - RemaningData)$

6. Return $L$

Figure 3 illustrates the flow chart of Algorithm 4:

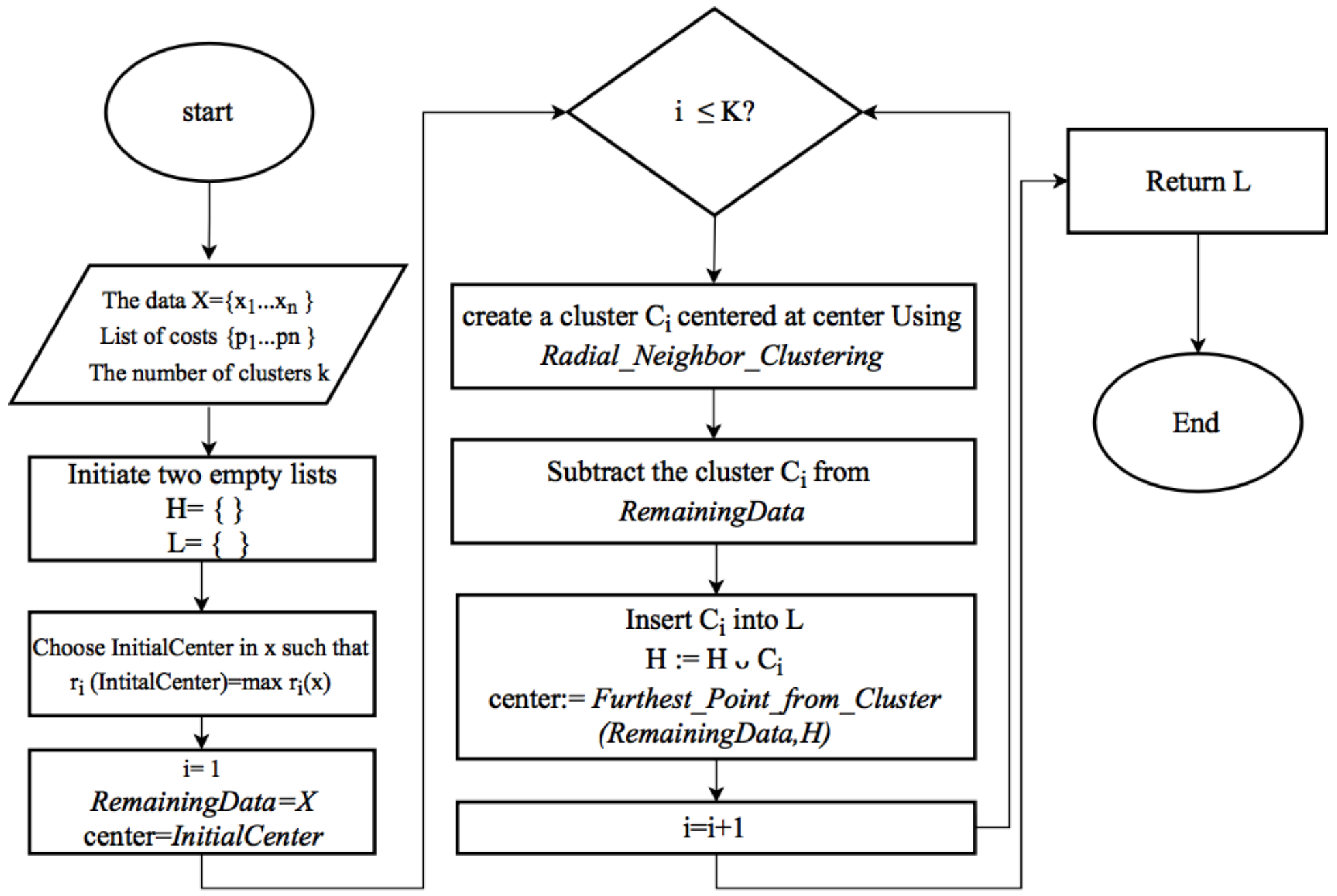

Figure 3: The flow chart of the ***Landark_Based_Radial_Clusterin***

Note that Algorithm 2 and 4 rely on the Algorithm 1 in forming the cluster at each stage. Recall that in Algorithm 1 we add the points to the center CN based on their closeness to the center. Namely, we order the points according to their distance to the center CN and then we added them in this order as long as we do not exceed the costs associated to the cluster at hand. For our purpose however, this is not always idea. Some other criterion such as the schedule of the project that the point represent and the cost of that project should also play a role in making the final cluster. We consider this issue in the next section.

**Consideration of the Schedule and the Cost Constrains in Forming the Clusters**

After choosing the center $CN$, the only criterion to add a point $x$ from the data $X$ a to the cluster C in Algorithm 1 is the distance between the point x and the center $CN$. The priority of adding a point $x$ to the cluster C depends only on is distance from the center $d(CN, x)$. We want to incorporate other criterion to the addition of a point a certain cluster. Our strategy to achieve this goal is simple and it goes as follows. We will assume that for each point $x$ we are given an integer-valued function $Y(x)$ representing the year at which is project is ideally executed. We call this function, the execution year function.

For each cluster center $CN$ with a cost p, we define two tolerance parameters $e_l$ and $e_h$, determined by the user. The purpose of the tolerance parameters is better explained after we explain our process. Next we run algorithm 1 on the three inputs $(X, CN, p)$, $(X, CN, p + e_h)$ and $(X, CN, p - e_l)$. The output on these inputs are respectively the clusters $C$, $C_h$ and $C_l$. Note that since these three clusters have the same center $CN$, they must be contained in each other as follows $C_l \subseteq C \subseteq C_h$. We will form a new cluster $C_F$ out of these clusters as follows. We start by adding all elements the small cluster $C_l$ to $C_F$. Then, we consider all elements in the set $C_h - C_l$ and we define a new order on these points as follows: we order the points in $C_h - C_l$ based on their execution function values. If two points from $C_h - C_l$ have the same year of execution then we put the ones, which have lower cost $f$ first in the order. After sorting we still have the same set $C_h - C_l$ but the points are not ordered in an order that is more desirable to us. In the final stage we add points from the ordered set $C_h - C_l$ to the $C_F$ as long as the total cost of the cluster of $C_F$ does not exceed $p$. Note that $C_F$ has the following properties. Every point that are very close to the center, those are the points in $C_l$, that we added first to the cluster $C_F$ is determined soley by its distance to the center. In other words, when the points are close to the center we prioritize its closeness over the other two parameters: year of execution and the cost of that point. This is determined by the parameter $e_l$. On the other hand, the parameter $e_l$ plays a role to determine the points that are a little further from the center. These are the points $C_h - C_l$, we add these points to the final cluster $C_F$ based on their year of execution and their costs. Points with sooner year of execution, or the ones that are passed due are given higher priority. Whenever two points in $C_h - C_l$ have the same year of execution we prioritize the one which has lower cost.

**Consideration of the Effect of Projects Reschedule on Performance and Cost**

Apparently, the proposed methodology will result in changing the initial M&R projects schedule developed by current PMS software products. This schedule change will subsequently change the cost of the maintenance according to the interest rate value as well as the pavement deterioration rate. In other words, a pavement segment $x$ that is predicted to have a PCI of "$I(x)$" at year $Y_i = Y_i(x)$, and assumed to cost a dollar amount of $f_{Y_i}(x)$ if repaired at that year $Y_i$, would not consume the same $f_{Y_i}(x)$ if it is rescheduled to be repaired in year $Y_{i\pm k}$. To overcome this challenge, unlimited budget scenarios are to be run separately, through the PMS software, at each year of the analysis period to develop a list $P_x$ for each project x, where, $P_x \coloneqq \{ f_{Y_1}(x), \dots, f_{Y_N}(x)\}$, and N is the number of years in the analysis period. When a project is moved from year $Y_i$ to year $Y_{i\pm k}$, the algorithm will utilize the corresponding $f_{Y_{i\pm k}}(x)$ from the developed list instead of the initial $f_{Y_i}(x)$. The following matrix illustrates the concept of the aforementioned unlimited budget scenarios runs:

| Scenario # | Year 1 | Year2 | ……. | Year N |
|---|---|---|---|---|
| 1 | $ Unlimited Budget | $0 Budget | ……. | $0 Budget |
| 2 | $0 Budget | $ Unlimited Budget | ……. | $0 Budget |
| ……. | ……. | ……. | ……. | ……. |
| N | $0 Budget | $0 Budget | ……. | $ Unlimited Budget |

## Analysis and Results

The introduced algorithm is not proposed to substitute any of the current PMS software products or optimization algorithms. Instead, the introduced methodology utilizes the current PMS software products' outputs (annual budget per M&R category and initial projects schedule) as inputs to feed the clustering algorithm. The algorithm utilizes these inputs to reschedule the projects such that the pavement segments with converged spatial coordinates will be repaired in the same fiscal year without compromising the allocated budget levels. The introduced methodology was utilized on two real-life examples of the City of Milton, GA and the City of Tyler Texas.

### Case study 1

To validate the proposed algorithm, a case study of approximately 800 projects in the City of Milton, GA was selected. First, the budget allocated annually for pavement maintenance was provided by the city engineers. Second, the PAVER software was utilized to create an initial five-year project planning. The map in Figure 4 presents the Global maintenance initial project planning obtained from the PAVER. As shown in Figure 4, projects assigned to the same fascial year are very scattered. The proposed algorithm was then applied to those global pavement maintenance projects using the initial PAVER recommended annual budget as constraints. Figure 5 presents the map of the global maintenance project planning after applying the proposed algorithm. As shown in Figure 5, the proposed project planning pavement segments with converged spatial coordinates are proposed to be repaired in the same fiscal. It also can be noticed that in the proposed planning projects are executed 1 year earlier due the residuals of the available budget, as projects initially allocated for the year 2021 were very few.

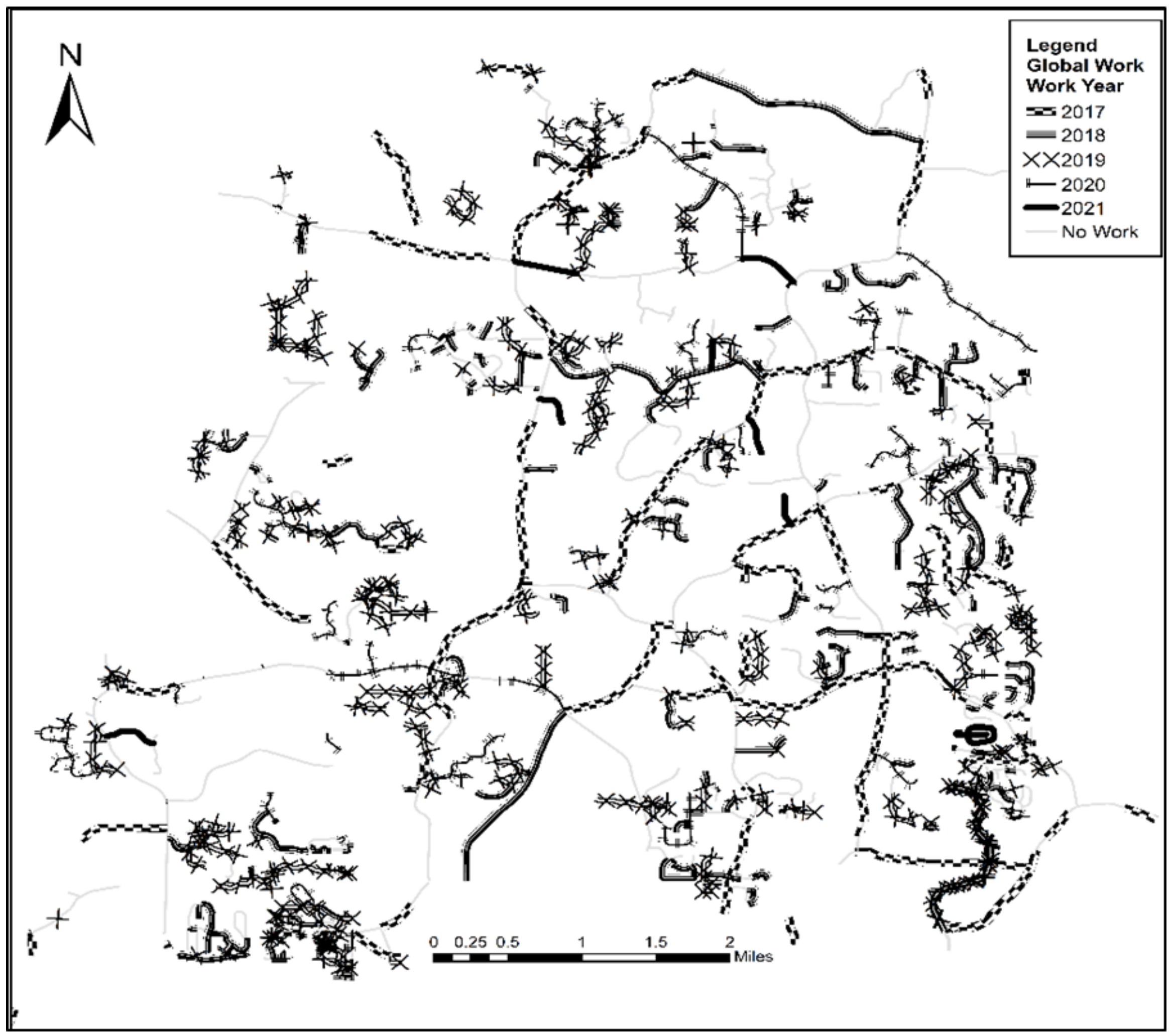

Figure 4: The Initial Global Pavement Maintenance Project Planning Map, City of Milton

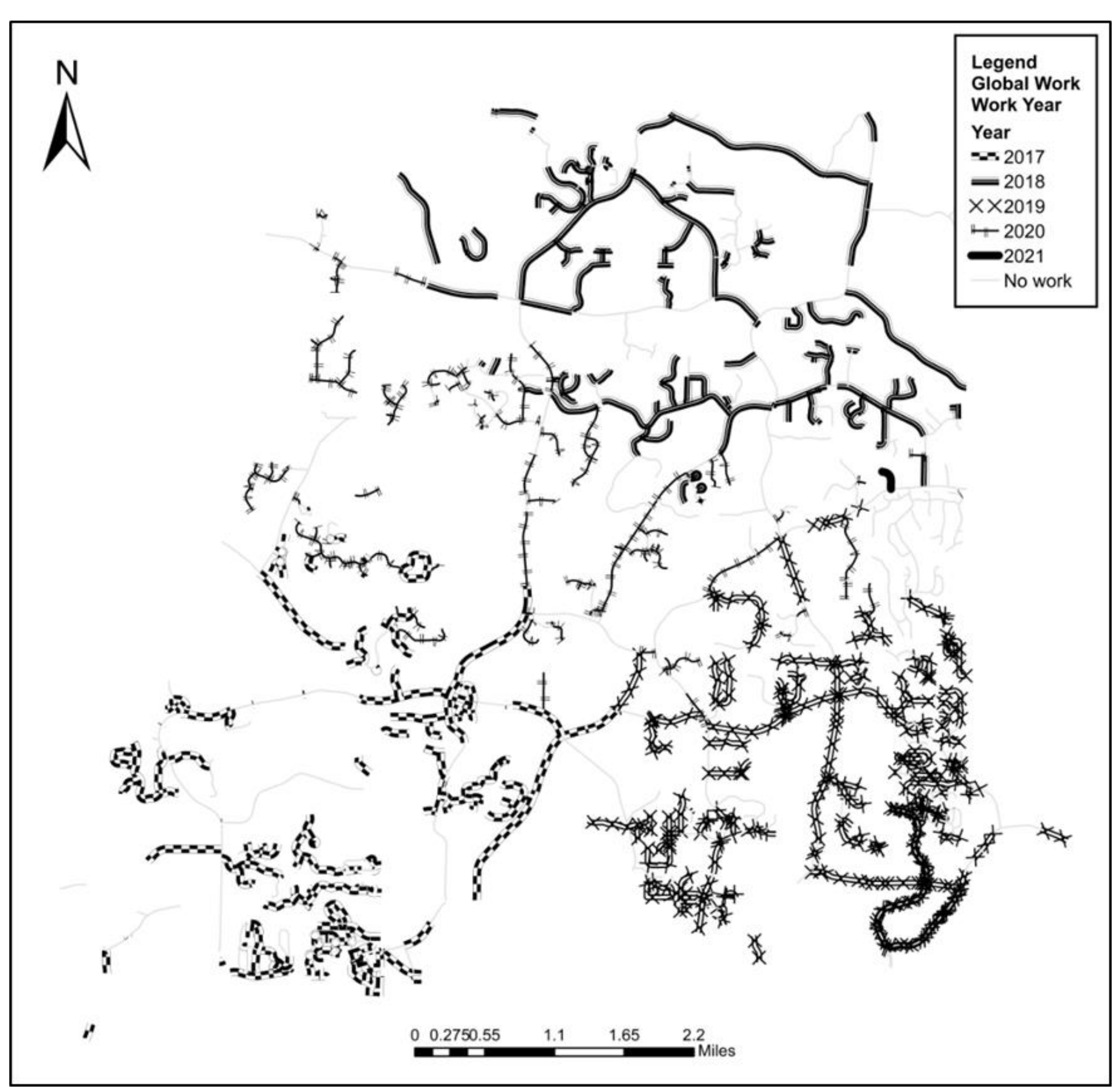

Figure 5: Proposed Global Pavement Maintenance Project Planning Map, City of Milton

**Case study 2**

**Applying the Algorithm**

For further validation of the proposed methodology, City of Tyler TX was selected as another case study. Approximately, 1000 pavement segments were included in the analysis. The Initial PAVER five-year project planning for major pavement maintenance projects is shown in Figure 6. The initial planning budget was used as constraints and the proposed algorithm was applied. Figure 7 presents the proposed project planning map after applying the algorithm.

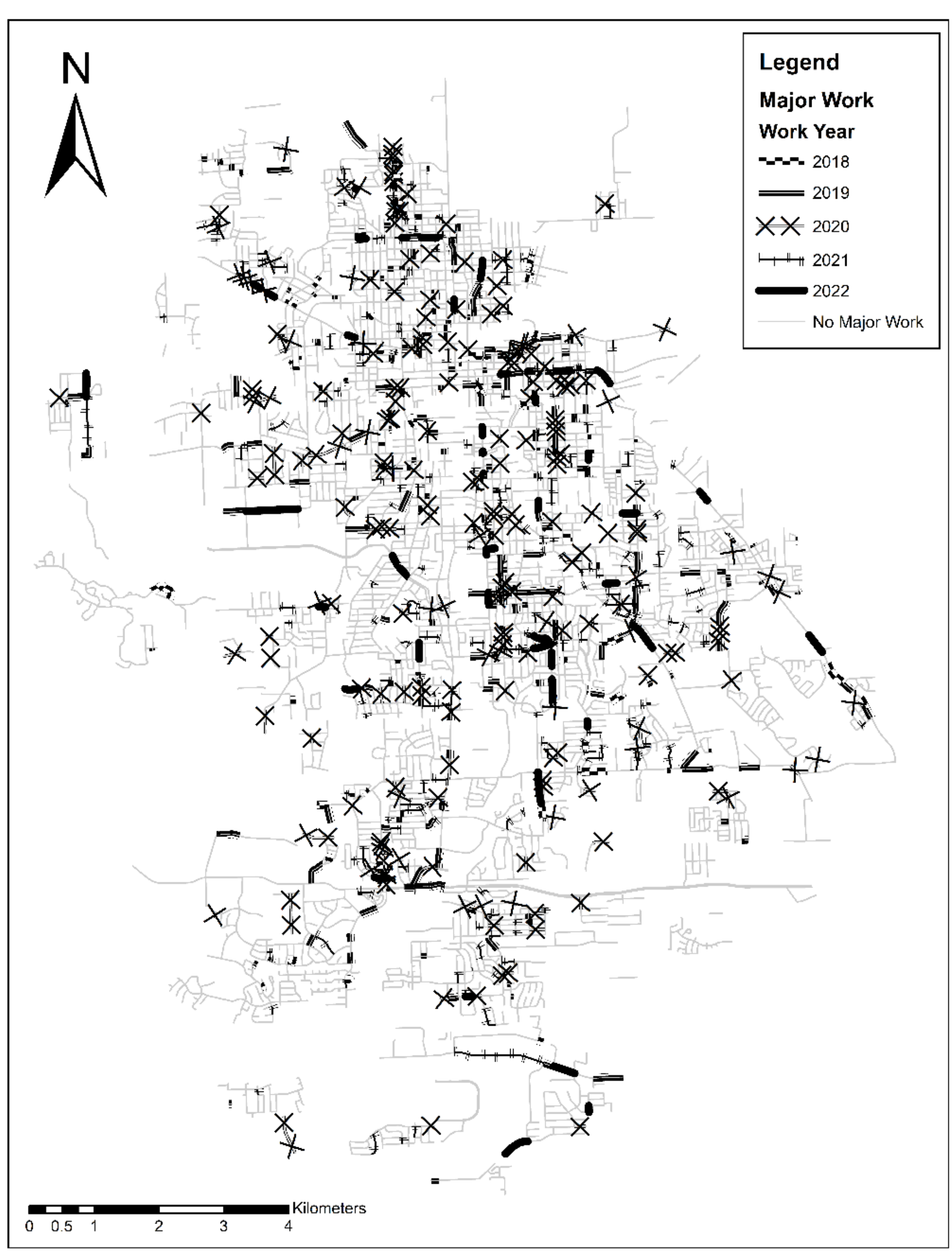

Figure 6: The Initial Global Pavement Maintenance Project Planning Map, City of Tyler

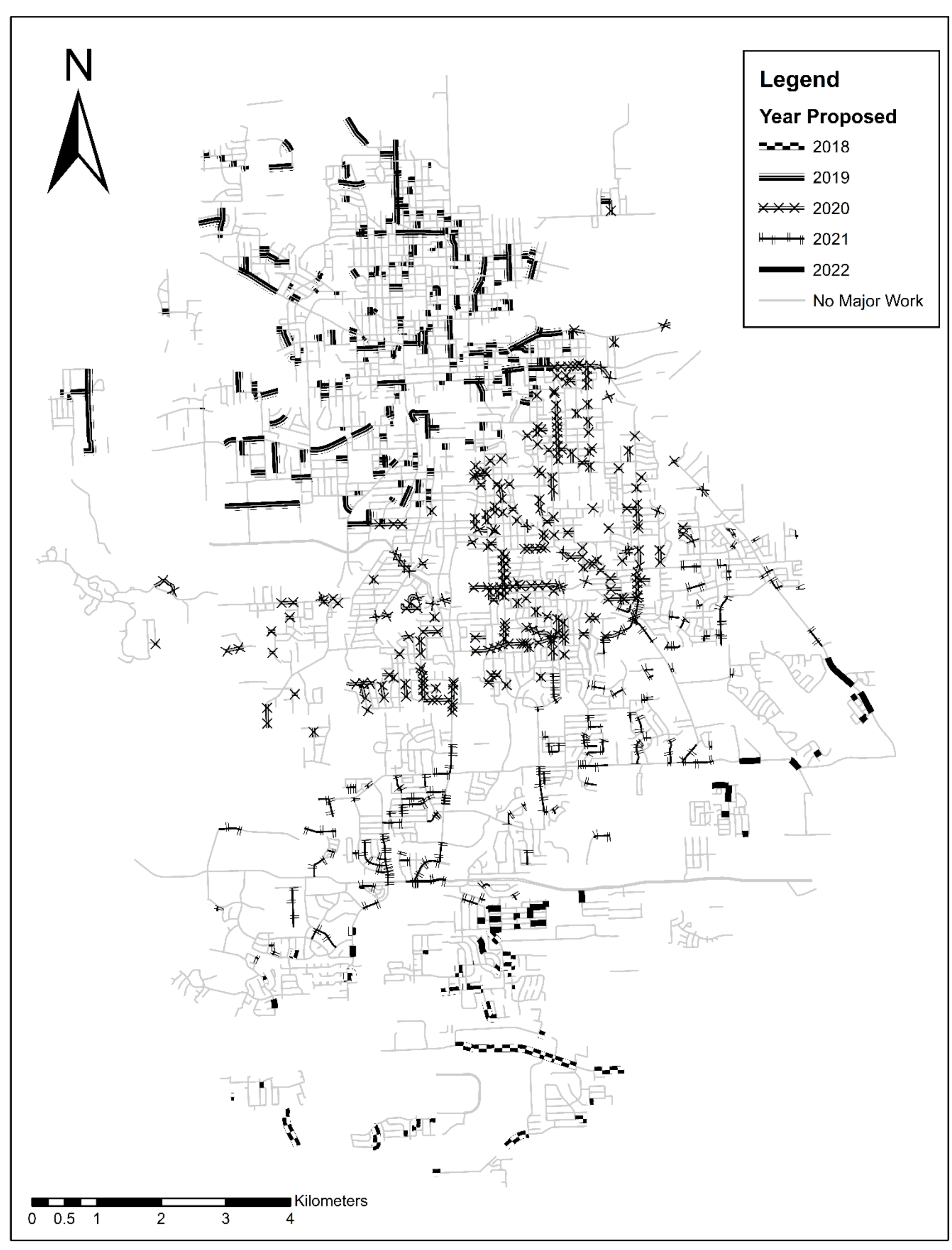

Figure 7: Proposed Global Pavement Maintenance Project Planning Map, City of Tyler

## Assess the Algorithm Influence on the Overall PCI

To confirm the proposed algorithm has a minimal effect on the City overall PCI, the PAVER software was utilized to calculate the predicted PCI throughout the project planning assuming both the initial schedule and the proposed one. As shown in Figure 8, the effect of applying the proposed algorithm on the repaired pavement segments overall PCI is very minimal. This is because the proposed algorithm utilizes the same budget distribution among the three aforementioned pavement maintenance categories recommended in the initial planning.

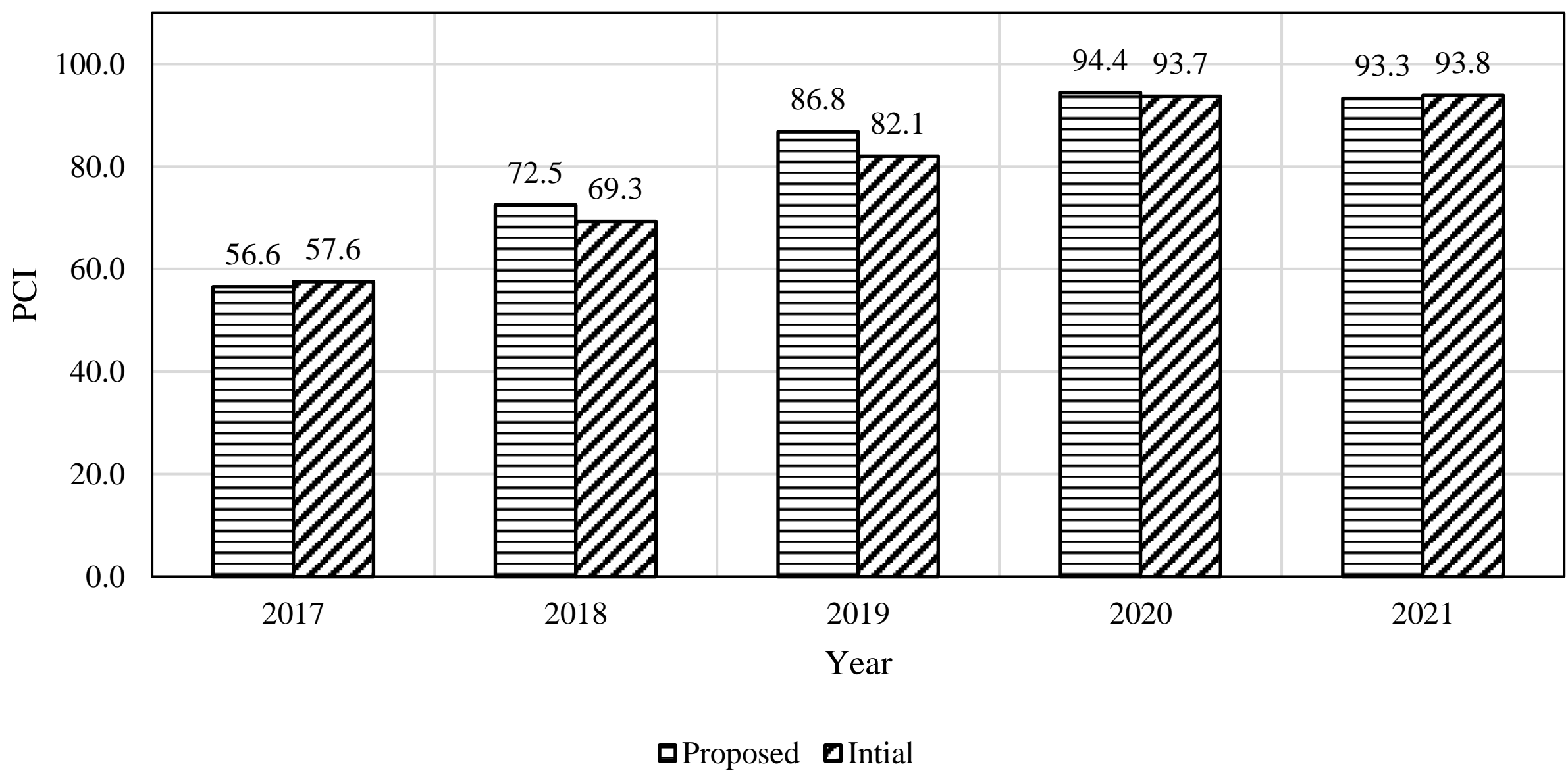

Figure 8: Influence of applying the proposed algorithm on the overall PCI

The study verifies regional clustering for pavement project priority and improves infrastructure management and decision analysis. First, multi-criteria weighted clustering is more practical than cost-condition models since it considers spatial closeness, budget, and scheduling. The program links optimization theory to pavement restoration logistics using these limits. This study restructures pavement maintenance and rehabilitation (M&R) schedules using geographical clustering and budgetary and temporal constraints, which current pavement management systems neglect. Unlike earlier methods that solely examine pavement quality, cost, and budget, this research provides a spatially conscious clustering algorithm that clusters geographically neighboring pavement segments for simultaneous treatment. Geo-coordinated scheduling reduces equipment deployment, enhances site monitoring, and promotes operational efficiency without hurting budget or network resilience. Adjustable weighted clustering parameters are utilized for cluster center location, project year, and cost. The rescheduling cost adjustment system adjusts treatment timing costs using degradation rates and time-dependent budget scenarios. Our method was adaptive, scalable, and had little impact on the network-wide Pavement Condition Index in Milton, GA and Tyler, TX. Study inspired PMS frameworks with spatial intelligence, automation, and data-driven decision support. Previously subjective and time-consuming, M&R schedule modification now automated. Automation helps transportation agencies manage thousands of pavement segments under real-world constraints by improving planning repeatability, transparency, and scalability. Model dynamic project rescheduling and time-dependent cost with annual budget simulation matrices to match project execution deferrals and advances to actual cost and PCI degradation. Rescheduling and geographical consistency were overlooked in previous clustering. It is also possible to integrate spatial intelligence into any Pavement Management System (PMS) without major design changes. This plug-and-play feature increases adoption. Milton, GA, and Tyler, TX, urban and semi-urban proof-of-concepts demonstrate the algorithm's network size and topological flexibility. The system may reduce bid prices and improve contractor coordination, saving money and enhancing project throughput, because it has no impact on network PCI and construction logistics. Finally, our study enhances spatially-aware PMS decision models for smart infrastructure, AI-enabled planning, and sustainable asset management. Intelligent clustering under limitations lays the framework for future research on integrating machine learning and optimization for infrastructure systems, leading to autonomous infrastructure planning platforms that optimize cost, condition, space, and time.

## Summary and Conclusions

This paper aimed at developing a clustering algorithm to combine pavement segments with converged spatial coordinates to be repaired in the same fiscal year without compromising the allocated budget levels or the overall pavement network condition. The proposed clustering algorithm naturally partitions the projects while considering simultaneously the budget constraints, the schedule, and spatial characteristics. The algorithm starts by finding the centroid for the clusters, then the clusters grow by gradually increasing the number of nearest neighbors around the centroids while taking into consideration the budget and the schedule constrains. Each cluster is finalized when its allocated budget is reached. The factors that form the clusters' constraints can be given weights to determine when the points join the cluster. Points with higher weights joint the cluster before points with lower weights. The developed algorithm was validated using 1,800 projects from two real-life examples of the City of Milton, GA and the City of Tyler, TX. According to the literature, analysis and results, the following conclusions can be drawn:

- Combining projects with converged spatial coordinates would result in minimizing the routing of crews, materials and other equipment among the construction sites and would provide better collaborations and communications between the construction teams;
- Utilizing the proposed methodology would assist the transportation agencies receiving lower bids from the contractors;
- The developed algorithm was successful in combining the M&R projects based on its spatial coordinates;
- The algorithm is capable of considering the annual budget limits as clustering constraints;
- Utilizing the proposed algorithm doesn’t affect the overall pavement network PCI.